\documentclass[aps,eqsecnum,preprint,floats,epsf,epsfig,nofootinbib,letter]{revtex4}
\usepackage{graphicx}
\usepackage{epsfig}

\usepackage{amssymb,epsfig}

\def\bea{\begin{eqnarray}}
\def\eea{\end{eqnarray}}

\def\pr{{Phys. Rev.}~}

\def\pl{{Phys. Lett.}~}

\def\lsim{ {\ \lower-1.2pt\vbox{\hbox{\rlap{$<$}\lower5pt\vbox{\hbox{$\sim$}
}}}\ } }
\def\gsim{ {\ \lower-1.2pt\vbox{\hbox{\rlap{$>$}\lower5pt\vbox{\hbox{$\sim$}
}}}\ } }

\begin{document}

\title{$B \to K_1 \gamma$ Decays in the Light-Cone QCD Sum Rules}

\author{\vspace*{0.5cm} \normalsize \sc Hisaki Hatanaka}
\author{\normalsize \sc Kwei-Chou Yang}

\affiliation{\vspace*{0.3cm} \normalsize\sl Department of Physics,
Chung Yuan Christian University, Chung-Li 320, Taiwan
\vspace*{2cm}}

\small
\begin{abstract}
\vskip0.5cm We present a detailed study of $B\to K_1(1270) \gamma$ and $B\to
K_1(1400) \gamma$ decays. Using the light-cone sum rule technique, we calculate
the $B\to K_{1A} (1^3P_1)$ and $B\to K_{1B} (1^1P_1)$ tensor form factors,
$T_1^{K_{1A}}(0)$ and $T_1^{K_{1B}}(0)$, where the contributions are included
up to the first order in $m_{K_1}/m_b$. We resolve the sign ambiguity of the
$K_1(1270)$--$K_1(1400)$ mixing angle  $\theta_{K_1}$  by defining the signs of
decay constants, $f_{K_{1A}}$ and $f_{K_{1B}}^\perp$. From the comparison of
the theoretical calculation and the data for decays $B\to K_1 \gamma$ and
$\tau^-\to K_1^-(1270)\nu_\tau$, we find that $\theta_{K_1}=-(34\pm 13)^\circ$
is favored. In contrast to $B\to K^* \gamma$, the hard-spectator contribution
suppresses the $B\to K_1(1270) \gamma$ and $B\to K_1(1400) \gamma$ branching
ratios slightly. The predicted branching ratios are in agreement with the Belle
measurement within the errors. We point out that a more precise measurement for
the ratio $R_{K_1}={\cal B}(B\to K_1(1400)\gamma)/{\cal B}(B\to
K_1(1270)\gamma)$ can offer a better determination for the $\theta_{K_1}$ and
consequently the theoretical uncertainties can be reduced.
\end{abstract}

\preprint{ }

\maketitle
\section{Introduction}
$b \to s \gamma$ decays contain rich phenomenologies relevant to the standard
model and new physics. Radiative $B$ decays involving a vector meson have been
observed by CLEO, Belle, and BaBar
\cite{Coan:1999kh,Nakao:2004th,Aubert:2004te}. Recently, the Belle
Collaboration has measured the $B\to K_1 \gamma$ decays for the first time
\cite{Abe:2004kr}:
\begin{eqnarray}
{\cal B}(B^-\to K_1^-(1270)\gamma)&=&(43\pm 9\pm 9)\times 10^{-6}~,
\\ {\cal B}(B^-\to K_1^-(1400)\gamma)&<&15\times 10^{-6}~,
\\ {\cal B}(\bar{B}^0\to \bar K_1^0(1270)\gamma)&<&58\times 10^{-6}~,
\\ {\cal B}(\bar{B}^0\to \bar K_1^0(1400)\gamma)&<&15\times 10^{-6}~,
\label{BELLE}
\end{eqnarray}
where $K_1$ is the orbitally excited (P-wave) axial-vector meson. The data
indicate that ${\cal B}(B\to K_1(1270) \gamma) \sim {\cal B}(B\to K^* \gamma)$
and ${\cal B}(B\to K_1(1270) \gamma) \gg {\cal B}(B\to K_1 (1400) \gamma)$.  It
is quite hard to explain the above-mentioned measurements using the existing
theoretical calculations
\cite{Altomari:1987pv,Veseli:1995bt,Safir:2001cd,Cheng:2004yj,Lee:2004ju,Kwon:2004ri}.
Therefore, these measurements represent a challenge for theory. The production
of the axial-vector mesons has been seen in the two-body hadronic $D$ decays
and in charmful $B$ decays \cite{PDG}. As for charmless hadronic $B$ decays,
$B^0\to a_1^\pm(1260)\pi^\mp$ are the first modes measured by $B$ factories
\cite{BaBara1pi,Bellea1pi}. The BaBar collaboration has recently reported the
observation of the decays $\bar B^0\to b_1^\pm\pi^\mp,b_1^+K^-$, $B^-\to
b_1^0\pi^-,b_1^0K^-,a_1^0\pi^-,a_1^-\pi^0$ \cite{Aubert:2007xd,:2007kp}, and
$\bar B^0\to K_1^-(1270)\pi^+,K_1^-(1400)\pi^+,a_1^+K^-$, $B^-\to a_1^-\bar
K^0,f_1(1285)K^-,f_1(1420)K^-$ \cite{Aubert:2007ds}. The related
phenomenologies have been studied in the literature
\cite{Yang1P1,Chen05,Nardulli05,Nardulli07,Calderon,Yang:2007sb,Cheng:2007mx}.

In this paper, we will focus on the study of the $B\to K_1 \gamma$ decays. The
physical states $K_1(1270)$ and $K_1(1400)$ are the mixtures of $1^3P_1$
($K_{1A}$) and $1^1P_1$ ($K_{1B}$) states. $K_{1A}$ and $K_{1B}$ are not mass
eigenstates and they can be mixed together due to the strange and nonstrange
light quark mass difference. Following the convention given in
Ref.~\cite{Suzuki:1993yc}, their relations can be written as
 \begin{eqnarray}
 \label{eq:mixing}
 |\bar K_1(1270)\rangle &=& |\bar K_{1A}\rangle\sin\theta_{K_1}+
  |\bar K_{1B}\rangle\cos\theta_{K_1}, \nonumber \\
 |\bar K_1(1400)\rangle &=& |\bar K_{1A}\rangle\cos\theta_{K_1} -
 |\bar K_{1B}\rangle\sin\theta_{K_1}.
 \end{eqnarray}
In Ref.~\cite{Suzuki:1993yc}, two possible solutions with two-fold ambiguity
$|\theta_{K_1}|\approx 33^\circ$ and $57^\circ$ were obtained. A similar
constraint $35^\circ\lesssim |\theta_{K_1}| \lesssim 55^\circ$ was found in
Ref. \cite{Burakovsky:1997ci}. From the data of $\tau\to K_1(1270)\nu_\tau$ and
$K_1(1400)\nu_\tau$ decays, the mixing angle is extracted to be $\pm37^\circ$
and $\pm 58^\circ$ in \cite{ChengDAP}. The sign ambiguity for $\theta_{K_1}$ is
due to the fact that one can add arbitrary phases to $|\bar K_{1A}\rangle$ and
$|\bar K_{1B}\rangle$. This sign ambiguity can be removed by fixing the signs
for $f_{K_{1A}}$ and $f_{K_{1B}}^\perp$, which do not vanish in the SU(3) limit
and are defined by
  \begin{eqnarray}\label{eq:k1a}
 \langle 0 |\bar \psi\gamma_\mu \gamma_5 s |\bar K_{1A}(P,\lambda)\rangle
 &=& -i \, f_{K_{1A}}\, m_{K_{1A}}\,\epsilon_\mu^{(\lambda)},
 \end{eqnarray}
\begin{eqnarray}\label{eq:k1b}
 \langle 0 |\bar \psi\sigma_{\mu\nu}s |\bar K_{1B}(P,\lambda)\rangle
 &=& i f_{K_{1B}}^\perp
 \,\epsilon_{\mu\nu\alpha\beta} \epsilon_{(\lambda)}^\alpha
 P^\beta,
 \end{eqnarray}
(with $\psi \equiv u$ {\rm or} $d$)  in the present paper. Following
Ref.~\cite{Yang:2007zt}, we adopt the convention: $f_{K_{1A}}>0$,
$f_{K_{1B}}^\perp>0$ and $\epsilon^{0123}=-1$. Thus, the signs of the $\bar B
\to \bar K_{1A,B}$ tensor form factors also depend on the definition mentioned
above. See also the discussions after Eq.~(\ref{eq:def-T1}).

In the quark model calculation, it was argued that the radiative $B$ decay
involving the $K_{1B}$ which is the pure $1^1P_1$ octet state is forbidden
because the effective operator $O_7$ is a spin-flip operator
\cite{Altomari:1987pv}. However, this is not true. Although, in the quark
model, the $1^1P_1$ meson is represented as a constituent quark-antiquark pair
with total spin $S=0$ and angular momentum $L=1$, a real hadron in QCD language
should be described in terms of a set of Fock states, for which each state with
the same quantum number as the hadron can be represented using light-cone
distribution amplitudes (LCDAs). In terms of LCDAs, the leading twist LCDAs of
the $\bar K_{1B}$ do not vanish, so that $\bar{B} \to \bar K_{1B}$ tensor form
factors are not zero. As a matter of fact, due to the G-parity, the
leading-twist LCDA $\Phi_\perp^{K_{1A}}$ ($\Phi_\parallel^{K_{1B}}$) of the
$\bar K_{1A}$ ($\bar K_{1B}$) meson defined by the nonlocal tensor current
(nonlocal axial-vector current) is antisymmetric under the exchange of $quark$
and $anti$-$quark$ momentum fractions in the SU(3) limit, whereas the
$\Phi_\parallel^{K_{1A}}$ ($\Phi_\perp^{K_{1B}}$) is symmetric
\cite{Yang:2005gk,Yang:2007zt}. The above properties were not well-recognized
in the previous light-cone (LC) sum rule calculation
\cite{Safir:2001cd,Lee:2006qj}.  In Ref.~\cite{Safir:2001cd}, the author used
only the ``symmetrically" asymptotic form for leading-twist distribution
amplitudes of the real states $K_1(1270)$ and $K_1(1400)$:
$\Phi_\perp^{K_1(1270)}(u)=\Phi_\perp^{K_1(1400)}(u)=6u\bar{u}$, in the LC sum
rule calculation. In Ref. \cite{Lee:2006qj}, only the $\bar{B} \to \bar K_{1B}$
tensor form factor $T_1^{K_{1B}}(0)$ (see Eq.~(\ref{eq:FF-1}) for the
definition) is computed. The correct forms of LCDAs for the axial-vector mesons
have been studied in details in Ref.~\cite{Yang:2007zt}. Using the LCDAs in
Ref.~\cite{Yang:2007zt}, $B\to K_1 \gamma$ decays have recently been
investigated in the perturbative QCD (PQCD) approach \cite{Wang:2007an}.

In this paper, making use of the LCDAs for the $\bar K_{1A}$ and $\bar K_{1B}$
in Ref.~\cite{Yang:2005gk,Yang:2007zt}, we study the $B\to K_1 \gamma$ decays.
We compute the relevant $\bar B\to \bar K_{1A}$ and $\bar K_{1B}$ tensor form
factors in the LC sum rule approach. The method of LC sum rules has been widely
used in the studies of nonperturbative processes, including weak baryon decays
\cite{Balitsky:1989ry}, heavy meson decays \cite{Chernyak:1990ag}, and heavy to
light transition form factors \cite{Belyaev:1993wp,Ball:1997rj,Ball:2004rg}. We
find that the $B\to K_1 \gamma$ data favor a negative $\theta_{K_1}$. The more
precise estimate can be made through the analysis for the $\tau^-\to
K_1^-(1270)\nu_\tau$ data. The predicted branching ratios for $B\to K_1(1270)
\gamma,  K_1(1400) \gamma$ are in agreement with the data within errors.

This paper is organized as follows. In Sec. \ref{sec:da}, the relevant
effective Hamiltonian is given. In Sec.~\ref{sec:Br-formula}, we provide the
definition of $\bar B\to \bar K_1$ tensor form factors and then gives the
formula for the $B\to K_1 \gamma$ branching ratios. In Sec.~\ref{sec:lcrs} we
derive the LC sum rules for the relevant tensor form factors, $T_{K_{1A}}$ and
$T_{K_{1B}}$. The numerical results and detailed analyses are given in
Sec.~\ref{sec:result}. We conclude in Sec.~\ref{sec:conclusion}. The relevant
expressions for two-parton and three-parton LCDAs are collected in Appendixes
\ref{app:2da-def} and \ref{app:3da-def}, respectively.

\section{The Effective Hamiltonian }\label{sec:da}

Neglecting doubly Cabibbo-suppressed contributions, the weak
effective Hamiltonian relevant to $b\to s\gamma$ is given by
\begin{equation}
{\cal H}_{\rm eff}(b\to s\gamma)=\frac{G_F}{\sqrt{2}}
\left\{V_{cb}V_{cs}^*
\left(c_1(\mu) O_1^c (\mu)+c_2(\mu) O_2^c(\mu)\right) -V_{tb}V_{ts}^*
 \sum_{i=3}^{8}c_i(\mu)O_i(\mu)\right\}~,
\end{equation}
where
\begin{eqnarray}
O_1^c &=& ({\overline{c}} b)_{V-A} ({\overline {s}} c)_{V-A},
 ~~ ~~~~~~~
O_2^c = ({\overline{c}_{\alpha}} b_{\beta})_{V-A}
 ({\overline {s}}_\beta c_\alpha)_{V-A}\,, \nonumber\\
O_3 &=& ({\overline{s}} b)_{V-A} \sum_q({\overline {q}} q)_{V-A},
 ~~~~
O_4 = ({\overline{s}_{\alpha}} b_{\beta})_{V-A} \sum_q({\overline
{q}}_\beta q_\alpha)_{V-A},
 \nonumber\\
O_5 &=& ({\overline{s}} b)_{V-A} \sum_q({\overline {q}} q)_{V+A}, ~~
~~ O_6 = ({\overline{s}_{\alpha}} b_{\beta})_{V-A} \sum_q({\overline
{q}}_\beta q_\alpha)_{V+A},\nonumber\\
O_7&=&\frac{em_b}{8\pi^2}\bar s_\alpha\sigma^{\mu\nu}
(1+\gamma_5)b_\alpha F_{\mu\nu}~, \nonumber\\
O_8&=&\frac{g_sm_b}{8\pi^2}\bar
s_\alpha\sigma^{\mu\nu}(1+\gamma_5)T^a_{\alpha\beta}b_\beta
G^a_{\mu\nu}~.
\end{eqnarray}
Here $\alpha, \beta$ are the $SU(3)$ color indices, $V\pm A$ correspond to
$\gamma^\mu (1 \pm \gamma^5)$, and we have neglected corrections due to the
$s$-quark mass. We will adopt the next-to-leading order (NLO) Wilson
coefficients computed in Ref.~\cite{Greub:1996tg}.


\section{The formula for the $B\to K_1 \gamma$ branching ratio}\label{sec:Br-formula}

The {\em penguin} form factors for $B \to K_1$  are defined as follows:
\begin{eqnarray}
 \left < \bar K_1 (p, \lambda)| \bar s \, \sigma_{\mu \nu} \gamma_5 q^\nu b
 | \bar B (p_B) \right > & = &
 2 T_1^{K_1} (q^2) \, \epsilon_{\mu\nu\rho\sigma} \,
 \epsilon_{(\lambda)}^{* \nu} \, p_B^\rho \, p^\sigma \,,
\label{eq:FF-1} \\
 \left < \bar K_1 (p, \lambda)  | \bar  s \, \sigma^{\mu \nu} q_\nu b
 |\bar B (p_B) \right > & = &
 -i T_2^{K_1} (q^2) \, [(m_B^2 - m_{K_1}^2)\, \epsilon_{(\lambda)}^{* \mu} -
       (\epsilon_{(\lambda)}^* q) \, (p + p_B)^\mu]
\qquad
\label{eq:FF-2} \\
& & - i T_3^{K_1} (q^2) \, (\epsilon_{(\lambda)}^* q) \, \left [
q^\mu - \frac{q^2}{m_B^2 - m_{K_1}^2} \, (p + p_B)^\mu \right ] ,
\nonumber
\end{eqnarray}
with
\begin{equation}
 T_1^{K_1}(0)  =  T_2^{K_1}(0). \label{eq:T1T2}
\end{equation}
where $\bar K_1$ can be $\bar K_{1A}$ or $\bar K_{1B}$ (or $\bar K_1(1270)$,
$\bar K_1(1400)$).

At the next-to-leading order of $\alpha_s$, the branching ratio can
be expressed as \cite{Ali:2001ez,Bosch:2001gv,Lee:2004ju}:
\begin{eqnarray}
{\cal B} (B \to K_1 \gamma) & = & \tau_B \, \Gamma (B \to K_1
\gamma)
\label{eq:br} \\
& = & \tau_B \,\frac{G_F^2 \alpha |V_{tb} V_{ts}^*|^2}{32 \pi^4} \,
m_{b, {pole}}^2 \, m_B^3 \, \left ( T_1^{K_1}(0) \right )^2
\left ( 1 - \frac{m_{K_1}^2}{m_B^2} \right )^3 \left | c^{(0){\rm
eff}}_7 +  A^{(1)} \right |^2 , \nonumber
\end{eqnarray}
where $m_{b,{pole}}$ is the pole mass of the $b$ quark, and $\alpha$ is the
electromagnetic fine structure constant. The effective coefficient $c_7^{(0)\rm
eff}$ in the naive dimensional regularization (NDR) scheme is defined by
$c_7^{(0)\rm eff}=c_7-\frac{1}{3}c_5-c_6$. $ A^{(1)}$ can be decomposed as
\begin{equation}
 A^{(1)} (\mu)  =   A_{C_7}^{(1)} (\mu) +
 A_{\rm ver}^{(1)} (\mu) +  A_{\rm sp}^{(1)K_1} (\mu_{\rm sp})~,
\label{eq:A1}
\end{equation}
where  $ A^{(1)}_{c_7}$, $ A^{(1)}_{\rm ver}$, which are the NLO
corrections due to the Wilson coefficient $c_7^{(0)\rm eff}$ and in the
$b \to s \gamma$ vertex, respectively, and $ A^{(1) K_1}_{\rm sp}$,
which is the hard-spectator correction, are given by
\begin{eqnarray}
 A_{c_7}^{(1)} (\mu) & = & \frac{\alpha_s (\mu)}{4 \pi} \,
c^{(1){\rm eff}}_7 (\mu) ,
\label{eq:A1tb-C7} \\
 A_{\rm ver}^{(1)} (\mu) & = & \frac{\alpha_s (\mu)}{4 \pi}
\left \{ \frac{32}{81} \left [ 13 c^{(0)}_1 (\mu) - 9 \,
c^{(0) {\rm eff}}_8 (\mu) \right ] \ln
\frac{\overline m_b}{\mu} \right .
\label{eq:A1tb-ver} \\
&  & +\left . \frac{4}{27} \left ( 33 - 2 \pi^2 + 6 \pi i \right ) c^{(0) {\rm
eff}}_8 (\mu) -\frac{16}{3}c_7^{(0)\rm eff}+ r_2 (z) \, c^{(0)}_1 (\mu) \right
\}, \qquad
\nonumber \\
 A_{\rm sp}^{(1)K_1} (\mu_{\rm sp}) & = &
 \frac{\pi\alpha_s (\mu_{\rm sp})C_F}{3 N_c} \,
 \frac{f_B f^\perp_{K_1} \lambda_B^{-1}}{m_B T_1^{K_1}(0)} \nonumber\\
 & & \times \left \{c^{(0){\rm eff}}_8 (\mu_{\rm sp})
   \langle  u^{-1} \rangle_\perp^{(K_1)}
 - c^{(0)}_1 (\mu_{\rm sp}) \langle \frac{\Delta i_5(z_0^{(c)},0,0)}
 {\bar u}\rangle_\perp \right \}.
 \nonumber\\ \label{eq:A1tb-sp}
\end{eqnarray}
Here $c_8^{\rm eff}=c_8+c_5$, $m_B/\lambda_B$ describes the first negative
moment of the $B$-meson distribution amplitude $\Phi_{B1}$
\cite{Beneke:2000ry,Bosch:2001gv}, and \begin{eqnarray}
 \langle  u^{-1}\rangle_\perp^{(K_1)}&\equiv&\int_0^1 du
 \frac{\Phi_\perp^{K_1}(u)}{ u}\,, \\
 \langle \frac{\Delta i_{5}(z_0^{(c)},0,0)}{\bar{u}}\rangle_\perp
&\equiv& \int_0^1 du \frac{\Delta i_{5}(z_0^{(c)},0,0)}{\bar{u}}
 \Phi^{K_1}_\perp(u) ,
 \label{eq:inverse-moment}
\end{eqnarray}
with $z = (\overline m_c/\overline m_b)^2$ and $z_0^{(c)} \simeq
m_B^2\bar{u}/\overline m_c^2$, where $\overline m_c\equiv \overline
m_c(\overline m_c)$ and $\overline m_b\equiv \overline m_b(\overline m_b)$ are
the $\overline{\rm MS}$ $c$- and $b$- quark masses, respectively. The detailed
definitions of the functions $r_2(z)$ and $\Delta i_5(z_0^{(c)},0,0)$ can be
found in Refs.~\cite{Greub:1996tg,Ali:2001ez}. In the numerical calculation, we
set the scale for the vertex corrections to be $\mu=\overline m_b$ and scale
for the spectator interactions to be $\mu_{sp}=\sqrt{\Lambda_h \overline m_b}$,
where $\Lambda_h\simeq$ 0.5~GeV corresponds to the hadronic scale.

\section{The light-cone sum rule for $T_1^{K_1}$}\label{sec:lcrs}

To calculate the form factor $T_1^{K_1}$, we consider the two-point correlation
function, which is sandwiched between the vacuum and transverse polarized $K_1$
meson,
\begin{eqnarray}
\lefteqn{i\int d^4x e^{iq x} \langle \bar K_1(P,\perp)|T [\bar
s(x)\sigma_{\mu\nu} b(x)\, j_B^\dagger(0)]|0\rangle }
\hspace*{1.0cm} \nonumber\\
  & & = -i\mathbb{A}(p_B^2,q^2)
\{\epsilon^{*(\perp)}_\mu (2P+q)_\nu - \epsilon^{*(\perp)}_\nu
(2P+q)_\mu\}
\nonumber\\
& & {}\ \  +i \mathbb{B}(p_B^2,q^2)\{\epsilon^{*(\perp)}_\mu q_\nu
- \epsilon^{*(\perp)}_\nu q_\mu\} + 2 i\mathbb{C} (p_B^2,q^2)
\,\frac{\epsilon^{*(\perp)} q}{m_B^2 -m_{K_1}^2}\,
\{P_\mu q_\nu - q_\mu P_\nu\},\nonumber\\ 
\label{eq:correlator-1}
\end{eqnarray}
where $j_B=i \bar \psi \gamma_5 b$ (with $\psi \equiv u$ {\rm or} $d$) is the
interpolating current for the $B$ meson, $p_B^2=(P+q)^2$, and $P$ the momentum
of the $K_1$ meson. Note that in this section $K_1 \equiv K_{1A}$ or $K_{1B}$.
$\mathbb{A}$ is the only relevant term in the present study, and at the hadron
level can be written in the form
\begin{equation}
  \mathbb{A}(p_B^2,q^2)=  T_1^{K_1}(q^2)\cdot \frac{1}{m^2_B-p_B^2} \cdot
        \frac{m_B^2 f_B}{m_b} + \cdots \,,
\end{equation}
where the dots denote contributions that have poles $p_B^2=m_{B^*}^2$ with
$m_{B^*}$ being the masses of the higher resonance $B^*$-mesons. To obtain the
result for $\mathbb{A}$, we have taken into account here the transverse
polarized $K_1$, instead of its longitudinal component, because for the
longitudinal $K_1$, $\mathbb{A}$ mixes with $\mathbb{B}$ and $\mathbb{C}$ for
an energetic $K_1$.

In a region of sufficiently large virtualities: $m_b^2 - p_B^2\gg \Lambda_{\rm
QCD}m_b$, with a small $q^2\geq 0$, the operator product expansion is
applicable in Eq.~(\ref{eq:correlator-1}), so that in QCD for an energetic
$K_1$ meson the correlation function in Eq.~(\ref{eq:correlator-1}) can be
represented in terms of the LCDAs of the $K_1$ meson:
\begin{eqnarray}
\lefteqn{i\int d^4x e^{iq x} \langle \bar K_1(P,\perp)|T [\bar
s(x)\sigma_{\mu\nu} b(x)\, j_B^\dagger(0)]|0\rangle }
\hspace*{1.0cm} \nonumber\\
  & & = \int_0^1 \frac{-i}{(q+ k)^2-m_b^2}
  {\rm Tr} \Big[\sigma_{\mu\nu} (\not\! q + \not\! k +m_b)\gamma_5 M_\perp^{K_1} \Big]
  \Bigg|_{k=uE n_-}  du\nonumber\\
  & & ~~~ + \frac{1}{4}\int_0^1 dv \int_0^1 D\underline{\alpha}
  \frac{ 2v E^2 (n_- q) \Big(f_{3K_1}^A{\cal A}(\underline{\alpha})
                    + f_{3K_1}^V{\cal V}(\underline{\alpha})\Big)
  {\rm Tr} (\sigma_{\mu\nu} \not\! \epsilon^*_{(\perp)} \not\! n_-)}
  { \Big\{m_b^2 - [q+(\alpha_1+\alpha_g v) E n_-]^2\Big\}^2 }
   \cr
 && ~~~ +{\cal O}\bigg(\frac{m_{K_1}^2}{E^2}\bigg) \,,~~~\label{eq:green-fn-2}
\end{eqnarray}
where $f_{3K_1}^A\sim {\cal O}(f_{K_1} m_{K_1})$, $f_{3K_1}^V\sim {\cal
O}(f_{K_1} m_{K_1})$, $E=|\vec{P}|$, $P^\mu =En_-^\mu + m_{K_1}^2 n_+^\mu/(4E)
\simeq E n_-^\mu$ with two light-like vectors satisfying $n_- n_+=2$ and
$n_-^2=n_+^2=0$. Here $E\sim m_b$ and we have assigned the momentum of the
$s$-quark in the $K_1$ meson to be
 \begin{eqnarray}
 k^\mu = u E n_-^\mu +k_\perp^\mu + \frac{k_\perp^2}{4 uE}n_+^\mu\,,
\end{eqnarray}
where $k_\perp$ is of order $\Lambda_{\rm QCD}$. In Eq.~({\ref{eq:green-fn-2}),
in calculating contributions due to the two-parton LCDAs of the $\bar K_1$ in
the momentum space, we have used the following substitution for the Fourier
transform of $\langle \bar K_1(P,\perp)|\bar s_{\alpha}(x) \,
\psi_{\delta}(0)|0\rangle$,
\begin{equation}
x^\mu \to -i \frac{\partial}{\partial k_{\mu}}\simeq -i \Bigg(
\frac{n_+^\mu}{2E}\frac{\partial}{\partial u} +
\frac{\partial}{\partial k_{\perp\, \mu}}\Bigg)\,,
\end{equation}
where the term of order $k_\perp^2$ is omitted.
Thus, we can obtain the light-cone transverse
projection operator $M^{K_1}_\perp$ of the $\bar K_1$ meson in the momentum space:
 \begin{eqnarray} \label{eq:MT}
M^{K_1}_\perp &=& i\frac{f^{\perp}_{K_1}}{4} E \Bigg\{ \not\!
\epsilon^{*(\lambda)}_\perp\not\! n_- \gamma_5 \,
   \Phi_\perp(u)\nonumber\\
&&  - \frac{f_{K_1}}{f_{K_1}^\perp}\frac{m_{K_1}}{E} \,\Bigg[ \not\!
\epsilon^{*(\lambda)}_\perp\gamma_5 \, g_\perp^{(a)}(u) -  \,
E\int_0^u dv\, \Phi_a(v)
       \not\! n_-\gamma_5 \, \epsilon^{*(\lambda)}_{\perp\mu} \,\frac{\partial}{\partial
         k_{\perp\mu}}
\cr && + \,i \varepsilon_{\mu\nu\rho\sigma} \,
       \gamma^\mu \epsilon_\perp^{*(\lambda)\nu} \,  n_-^\rho
         \left(n_+^\sigma \,{g_\perp^{(v)\prime}(u)\over 8}-
          E\,\frac{g_\perp^{(v)}(u)}{4} \, \frac{\partial}{\partial
         k_\perp{}_\sigma}\right)
 \Bigg]
 \, \Bigg|_{k=up} \cr
 && +{\cal O}\bigg(\frac{m_{K_1}^2}{E^2}\bigg) \Bigg\}\,,
\end{eqnarray}
where $\Phi_a\equiv \Phi_\parallel -g_\perp^{(a)}$ and the detailed definitions
for the relevant two-parton LCDAs are collected in Appendix~\ref{app:2da-def}.
A similar discussion for the vector meson projection operators can be found in
Ref.~\cite{Beneke:2000wa}. From the expansion of the transverse projection
operator, one can find that contributions arising from $\Phi_a,
g_\perp^{(v)\prime}$, and $g_\perp^{(v)}$ are suppressed by $m_{K_1}/E$ as
compared with that from $\Phi_\perp$. Note that in Eq.~(\ref{eq:green-fn-2})
the derivative with respect to the transverse momentum acts on the hard
scattering amplitude before the collinear approximation is taken. The
three-parton chiral-even distribution amplitudes of twist-3, ${\cal
A}(\underline{\alpha})$ and ${\cal V}(\underline{\alpha})$, together with their
decay constants, $f_{3K_1}^A$ and $f_{3K_1}^V$, are defined by
\begin{eqnarray}
 \langle \bar K_1(P,\lambda)|\bar s(x) \gamma_\alpha\gamma_5 g_s
G_{\mu\nu}(vx) \psi(0)|0\rangle &=&
 p_\alpha[p_\nu\epsilon^{*(\lambda)}_{\perp\mu}
  -p_\mu \epsilon^{*(\lambda)}_{\perp\nu}]
      f_{3K_1}^A{\cal A}(v,-px) \nonumber\\
 & & +\cdots \,,
      \label{eq:3plcda-t3-1}\\
 \langle \bar K_1(P,\lambda)|\bar s(x) \gamma_\alpha
         g_s\widetilde G_{\mu\nu}(vx) \psi(0)|0\rangle &=&
 i  p_\alpha[p_\mu \epsilon^{*(\lambda)}_{\perp\nu}-p_\nu
\epsilon^{*(\lambda)}_{\perp\mu}]
      f_{3K_1}^V{\cal V}(v,-px)\nonumber\\
 &  & +\cdots
      \,,\label{eq:3plcda-t3-2}
\end{eqnarray}
where we have set $p_\mu=P_\mu-m_{K_1}^2 \bar z_\mu/(2 P \bar z)$ with
$$\bar z_\mu=x_\mu -\frac{P_\mu}{m_{K_1}^2}\Bigg\{xP- \Big[(xP)^2-x^2 m_{K_1}^2\Big]^{1/2}\Bigg\}\,.$$
Here the ellipses stand for terms of twist higher than three, the
following shorthand notations are used:
\begin{equation}
   {\cal A}(v,-px) \equiv  \int {\cal D}\underline{\alpha}
   \,e^{ipx(\alpha_{1}+v\alpha_g)}{\cal A}
   (\underline{\alpha}), \label{eq:short}
\end{equation}
etc., and the integration measure is defined as
\begin{equation}
 \int {\cal D}\underline{\alpha} \equiv \int_0^1 d\alpha_{1}
  \int_0^1 d\alpha_{2}\int_0^1 d\alpha_g \,\delta(1-\sum \alpha_i),
\label{eq:measure}
\end{equation}
with $\alpha_1, \alpha_2, \alpha_g$ being the momentum fractions carried by the
$s$ quark, $\bar\psi (\equiv \bar u\ {\rm or}\ \bar d)$ quark, and gluon,
respectively. At the quark-gluon level, after performing the integration of
Eq.~(\ref{eq:green-fn-2}), the result for $\mathbb{A}^{\rm QCD}$ reads (with
$\bar u=1-u$)
\begin{eqnarray}
  \mathbb{A}^{\rm QCD}&=& - \frac{m_b f_{K_1}^\perp}{2} \int_0^1 du \Bigg\{
  \frac{1}
  {m_b^2 -up_B^2 -\bar u q^2} \nonumber\\
 & & \ \ \
 \times\Bigg[ \Phi^\perp(u) -
  \frac{ m_{K_1}f_{K_1} }{m_b f_{K_1}^\perp }
  \Bigg(ug^{(a)}_\perp(u) + \Phi_a(u) + \frac{g^{(v)}_\perp(u)}{4}
  -\frac{g^{(v)\prime}_\perp(u)}{4}\frac{p_B^2+q^2}{p_B^2 -q^2}\Bigg)\Bigg] \nonumber\\
  & & \ \ \ \ -
  \frac{ m_{K_1}f_{K_1} }{4 m_b f_{K_1}^\perp }
   \frac{(m_b^2+q^2)}{(m_b^2-up_B^2 -\bar u q^2)^2}g^{(v)}_\perp(u)
   \Bigg\}\nonumber\\
   & & - \int_0^1 v dv \int_0^1 D\underline{\alpha}
  \frac{f_{3K_1}^A{\cal A}(\underline{\alpha})
                + f_{3K_1}^V{\cal V}(\underline{\alpha})}
   {2(\alpha_1 +v \alpha_g)}  \left[\frac{1}{m_b^2 - (\alpha_1+v \alpha_g)(p_B^2 -q^2) -q^2}\right.\nonumber\\
 && \ \ \  \left. - \frac{m_b^2-q^2}{[m_b^2 - (\alpha_1+v \alpha_g)(p_B^2 -q^2) -q^2 ]^2 } \right]\,.
\end{eqnarray}
We have given the results of $\mathbb{A}$ from the hadron and quark-gluon
points of view, respectively. Thus, the contribution due to the lowest-lying
$K_1$ meson can be further approximated with the help of quark-hadron duality:
\begin{equation}
  T_1^{K_1}(q^2)\cdot \frac{1}{m^2_B-p_B^2} \cdot
        \frac{m_B^2 f_B}{m_b} = \frac{1}{\pi}\int_{m_b^2}^{s_0}
        \frac{{\rm Im}\mathbb{A}^{\rm QCD}(s,q^2)}{s-p_B^2}ds\,,
\end{equation}
where $s_0$ is the excited state threshold. After applying the Borel transform
$p_B^2 \to M^2$  to the above equation, we obtain
\begin{eqnarray}
 T_1^{K_1}(q^2) = \frac{m_b}{m_B^2 f_B}   e^{-m_B^2/M^2} \frac{1}{\pi}\int_{m_b^2}^{s_0}
       e^{s/M^2} {\rm Im}\mathbb{A}^{\rm QCD}(s,q^2) ds\,.
 \end{eqnarray}
Finally, the light-cone sum rule for $T_1^{K_1}$ reads
\begin{eqnarray}\label{eq:T1-SR}
 T_1^{K_1}(q^2) =
  & - & \frac{m_b^2 f_{K_1}^\perp }{2 m_B^2 f_B}
  e^{(m_B^2 -m_b^2)/M^2} \int_0^1 du \Bigg\{ \frac{1}{u}
  e^{\bar u(q^2 -m_b^2)/(uM^2)} \theta[c(u,s_0)] \Bigg[\Phi^\perp(u)\nonumber\\
& &  \ \
 -\frac{ m_{K_1}f_{K_1} }{m_b f_{K_1}^\perp } \Big(u g^{(a)}_\perp(u) + \Phi_a(u)
  + \frac{g^{(v)}_\perp(u)}{4}
  -\frac{g^{(v)\prime}_\perp(u)}{4}\frac{m_b^2 +(u-\bar u) q^2}{m_b^2 -q^2} \Big) \Bigg] \nonumber\\
& &  - \frac{1}{u} e^{\bar u(q^2 -m_b^2)/(uM^2)}
 \frac{1}{4} \frac{m_{K_1} f_{K_1}}{m_b f_{K_1}^\perp }
 (m_b^2+q^2)g^{(v)}_\perp(u)
 \Bigg( \frac{\theta[c(u,s_0)]}{uM^2} +  \delta[c(u,s_0)] \Bigg) \nonumber\\
 & & -\frac{m_{K_1} f_{K_1}}{m_b f_{K_1}^\perp }
   \frac{g^{(v)\prime}_\perp(u)}{2}\frac{q^2}{m_b^2 -q^2} e^{(m_b^2-q^2)/M^2}\Bigg\}\nonumber\\
 & - &   \frac{m_b}{2m_B^2 f_B}   e^{(m_B^2 -m_b^2)/M^2} \int_0^1 v dv \int_0^1 D\underline{\alpha}
  \frac{f_{3K_1}^A{\cal A}(\underline{\alpha})
                + f_{3K_1}^V{\cal V}(\underline{\alpha})}
   {(\alpha_1 +v \alpha_g)^2} \nonumber\\
  & & \times e^{(1-\alpha_1-v\alpha_g)(q^2 -m_b^2)/[(\alpha_1+v\alpha_g)M^2]}
        \Bigg\{\theta[c(\alpha_1+v\alpha_g,s_0)]\nonumber\\
  & & \ \  - (m_b^2-q^2) \Bigg( \frac{\theta[c(\alpha_1+v\alpha_g,s_0)]}{(\alpha_1+v\alpha_g) M^2}
  +  \delta[c(\alpha_1+v\alpha_g,s_0)] \Bigg) \Bigg\},
\end{eqnarray}
where $c(u,s_0)=us_0 -m_b^2 + (1-u) q^2$ and $\theta[\cdots]$ is the step
function. Note that here $f_{K_{1A}}^\perp$ is chosen to be $f_{K_{1A}}$, while
$f_{K_{1B}}$ is adopted to be $f_{K_{1B}}^\perp(1~{\rm GeV})$. (See
Eq.~(\ref{app:decay}) and related discussions.)

\section{Results}\label{sec:result}

\subsection{$T_1^{K_{1A}}$ and $T_1^{K_{1B}}$ LCSR results and $B \to K_1 \gamma$ branching ratios}\label{subsec:result-1}
\begin{table}[t]
\begin{center}
{\tabcolsep=0.461cm\begin{tabular}{|c|c|c|c|c|c|} \hline\hline
\multicolumn{6}{|c|}{Running quark masses (GeV), pole $b$-quark mass (GeV), and couplings } \\
 \hline   $\overline m_c(\overline m_c)$ & $m_s(2\,\mbox{GeV})$
       & $\overline m_b(\overline m_b)$ & $m_{b,pole}$ & $\alpha_s(m_Z)$ & $\alpha$ \\
\hline
 $1.25\pm 0.10$ & $0.09\pm 0.01$ & $4.25\pm 0.15$ & $4.90\pm0.05$ & $0.1176$ & 1/137\\
\hline
\end{tabular}}
{\tabcolsep=1.113cm\begin{tabular}{|c|c|c|} \hline
\multicolumn{3}{|c|}{CKM matrix elements and the moment of the $B$ distribution amplitude} \\
\hline  $|V_{cs}|$ &  $|V_{cb}|$  & $\lambda_B$ \\
\hline $0.957\pm 0.095$ & $(41.6\pm 0.6)\times 10^{-3}$ & $(0.35\pm0.15)$ GeV \\
\hline
\end{tabular}}
{\tabcolsep=0.64cm\begin{tabular}{|c|c|c|c|c|} \hline
\multicolumn{5}{|c|}{Masses (GeV) and decay constants (MeV) for mesons} \\
\hline
$m_{K_{1A}}$ & $m_{K_{1B}}$ & $f_{K_{1A}}$ & $f_{K_{1B}}^\perp$(1~GeV) & $f_{B}$ \\
\hline
$1.31\pm 0.06$ & $1.34\pm 0.08$ & $250\pm 13$ & $190\pm 10$ & $190\pm 10$\\
\hline
\end{tabular}}
{\tabcolsep=0.3cm\begin{tabular}{|c|c|c|c|c|} \hline
\multicolumn{5}{|c|}{Gegenbaur moments for the $K_{1A}$ meson at scales 1 GeV and $2.2$~GeV (in parentheses)}\\
\hline
$ ~~~~~~a_1^{\parallel, K_{1A}}~~~~~$ & $a_2^{\parallel, K_{1A}}$ & ~~~~~$a_0^{\perp, K_{1A}}$~~~~ & $a_1^{\perp, K_{1A}}$ & $a_2^{\perp, K_{1A}}$\\
\hline
$-0.30^{+0.26}_{-0.00}$ & $-0.05\pm 0.03$ & $0.26^{+0.03}_{-0.22}$ & $-1.08\pm 0.48$ & $0.02\pm 0.20$\\
($-0.24^{+0.21}_{-0.00}$) & ($-0.04\pm 0.02$) & ($0.24^{+0.03}_{-0.21}$) & ($-0.84\pm 0.37$) & ($0.01\pm 0.15$)\\
\hline
\end{tabular}}
{\tabcolsep=0.29cm\begin{tabular}{|c|c|c|c|c|} \hline
\multicolumn{5}{|c|}{Gegenbaur moments for the $K_{1B}$ meson at scales 1 GeV and $2.2$~GeV (in parentheses)} \\
\hline
$ a_0^{\parallel, K_{1B}}$ & $a_1^{\parallel, K_{1B}}$ & ~~~~~$a_2^{\parallel, K_{1B}}$~~~~~ & ~~~~~$a_1^{\perp, K_{1B}}$~~~~~ & $a_2^{\perp, K_{1B}}$\\
\hline
$-0.15\pm 0.15$ & $-1.95\pm 0.45$ & $0.09^{+0.16}_{-0.18}$ & $0.30^{+0.00}_{-0.31}$ & $-0.02\pm 0.22$\\
($-0.15\pm 0.15$) & ($-1.56\pm 0.36$) & ($0.06^{+0.11}_{-0.13}$) & ($0.25^{+0.00}_{-0.26}$) & ($-0.02\pm 0.17$)\\
\hline
\end{tabular}}
{\small{\tabcolsep=0.19cm\begin{tabular}{|c|c|c|c|c|c|} \hline
\multicolumn{6}{|c|}{Parameters of twist-3 3-parton LCDAs of the $K_{1A}$ meson at the scale 2.2 GeV } \\
\hline
 $f^V_{3,K_{1A}}$ (in GeV$^2$) & $\omega_{K_{1A}}^V$ & $\sigma_{K_{1A}}^V$ & $f^A_{3,K_{1A}}$ (in GeV$^2$)& $\lambda_{K_{1A}}^A$ & $\sigma_{K_{1A}}^A$ \\
\hline
 $0.0034\pm 0.0018$ &  $-3.1  \pm 1.1$ & $-0.13\pm 0.16$ &
 $0.0014\pm 0.0007$ &  $0.70  \pm 0.46$ & $2.4\pm 2.0$\\
\hline
\end{tabular}}
}
{\small{\tabcolsep=0.15cm\begin{tabular}{|c|c|c|c|c|c|} \hline
\multicolumn{6}{|c|}{Parameters of twist-3 3-parton LCDAs of the $K_{1B}$ meson at the scale 2.2 GeV } \\
\hline
 $f^V_{3,K_{1B}}$ (in GeV$^2$) & $\lambda_{K_{1B}}^V$ & $\sigma_{K_{1B}}^V$ & $f^A_{3,K_{1B}}$ (in GeV$^2$)& $\omega_{K_{1B}}^A$ & $\sigma_{K_{1B}}^A$ \\
\hline
 $0.0029 \pm 0.0012$ & $0.09  \pm 0.24$ & $ 0.31\pm 0.68$ &
 $-0.0041\pm 0.0018$ & $-1.7  \pm 0.4$ &  $-0.05\pm 0.04$\\
\hline
\end{tabular}}}
\end{center}
\centerline{\parbox{14cm}{\caption{\label{tab:inputs} Summary of
input parameters \cite{PDG,Beneke:2000ry,Yang:2007zt}.}}} \vspace{0.1cm}
\end{table}

Parameters relevant to the present study are collected in
Table~\ref{tab:inputs}. We first analyze the $T_1(0)$ sum rules numerically.
The pole $b$ quark mass is  adopted in the LC sum rule. The $f_{K_1}^\perp$ and
parameters appearing in the distribution amplitudes are evaluated at the
factorization scale $\mu_f=\sqrt{m_B^2-m_{b,pole}^2}$. On the other hand, the
form factor $T_1(0)$ depends on the renormalization scale of the effective
Hamiltonian, for which the scale is set to be $\overline m_b(\overline m_b)$.
The working Borel window is $7.0$~GeV$^2< M^2 < 13.0$~GeV$^2$, where the
correction originating from higher resonance states amounts to 15\% to 35\%. We
do not include the contributions of the twist-4 LCDAs and 3-parton twist-3
chiral-even LCDAs in the light-cone sum rule since these corrections to
light-cone expansion series is of order $(m_{K_1}/m_b)^2$ and might be
negligible. The excited state threshold $s_0$ can be determined when the most
stable plateau of the LC sum rule result is obtained within the Borel window.
We find that the corresponding threshold $s_0$ lies in the interval $32 \sim
36$~GeV$^2$.

Two remarks are in order. First, we have consistently used $f_B=190\pm 10$~MeV
in all numerical analysis. In the literature, it was {\it assumed} that the
theoretical errors due to the radiative corrections in the form factor sum
rules can be canceled if one adopts the $f_B$ sum rule result with the same
order of $\alpha_s$-corrections in the calculation
\cite{Ball:1997rj,Ball:2004rg}. Nevertheless, the resulting sum rule result for
$T_1^{BK*}(0)$ seems to be significantly larger than the estimate extracted
from the data \cite{Ali:2001ez}, although the sum rule result can be improved
by including $\alpha_s$-corrections \cite{Ball:2004rg}. We have checked that
using the physical value of $f_B$, that we adopt here, in the $T_1^{BK^*}(0)$
LC sum rule with the same order in $\alpha_s$ and $m_{K_1}/m_b$, we get
$T_1^{BK^*}(0)\approx 0.25^{+0.03}_{-0.02}$ which is in good agreement with the
result constrained by the data \cite{Ali:2001ez,Ball:2006eu}. Extracting from
the data, the current estimation is $T_1^{BK^*}(0)=0.267\pm 0.018$
\cite{Ball:2006eu}. The lattice QCD result is $T_1^{BK^*}(0)=0.24\pm
0.03^{+0.04}_{-0.01}$ \cite{Becirevic:2006nm}. Therefore, although the
radiative corrections can be important in the form factor sum rule
calculations, its effects are significantly reduced and may be negligible in
the present analysis. Second, $a_1^{\parallel, K_{1A}}, a_0^{\perp, K_{1A}},
a_2^{\perp, K_{1A}}, a_0^{\parallel, K_{1B}},  a_2^{\parallel, K_{1B}}$, and
$a_1^{\perp, K_{1B}}$ are G-parity violating Gegenbaur moments, which vanish in
the SU(3) limit. Using the QCD sum rules, the relation $a_0^{\perp, K_{1A}}
+(0.59\pm 0.15)a_0^{\parallel, K_{1B}} =0.17\pm 0.11$ was obtained, instead of
their individual values \cite{Yang:2007zt}. It will be seen later that due to
the data for ${\cal B}(B\to K_1(1270) \gamma) \gg {\cal B}(B\to K_1(1400)
\gamma)$ and for $\tau^-\to K_1^-(1270)\nu_\tau$, $\theta_{K_{1}}$ and
$a_0^{\parallel, K_{1B}}$ should be negative. Here we further make  reasonable
assumptions that $|a_0^{\parallel, K_{1B}} f_{K_{1B}}| \leq 30\% \times
f_{K_{1B}}^\perp$ and $|a_0^{\perp, K_{1A}} f_{K_{1A}}^\perp| (1~{\rm GeV})\leq
30\% \times f_{K_{1A}}$ to account for the possible SU(3) breaking effect,
i.e., we assume G-parity correction is roughly less than 30\%. (See
Eqs.~(\ref{eq:k1-1v})-(\ref{eq:k1-2t}) for the detailed definitions of
parameters.) Finally, we arrive at $a_0^{\parallel, K_{1B}}=-0.15\pm 0.15$ and
$a_0^{\perp, K_{1A}}=0.26^{+0.04}_{-0.22}$. As shown in Table~\ref{tab:inputs},
once these two parameters are determined, the remaining G-parity violating
Gegenbaur moments are thus updated according to the relations given in
Eq.~(141) in Ref.~\cite{Yang:2007zt}.

To illustrate the qualities and uncertainties of the sum rules, we plot the
results for $T_1^{K_{1A}}(0)$ and $T_1^{K_{1B}}(0)$ as functions of $M^2$ in
Fig.~\ref{fig:T1}. We obtain
\begin{eqnarray}\label{eq:T1-result}
 T_1^{K_{1A}}(0)&=&\, ~~~~0.31^{+0.06+0.01+0.06}_{-0.04-0.01-0.03}\,, \nonumber\\
 T_1^{K_{1B}}(0)&=& -\left(0.25_{-0.02-0.01-0.07}^{+0.03+0.01+0.05}\right),
 \end{eqnarray}
where the first, second, and third error bars come from the variations of
$m_{b,pole}$, $f_B$, and the remaining parameters, respectively. The third
errors are mainly due to the G-parity violating Gegenbaur moments of the
leading twist LCDAs. Corrections arising from the three-parton LCDAs are less
than 3\%.
\begin{figure}[tb]
 \centerline{{\epsfxsize=6.7cm \epsffile{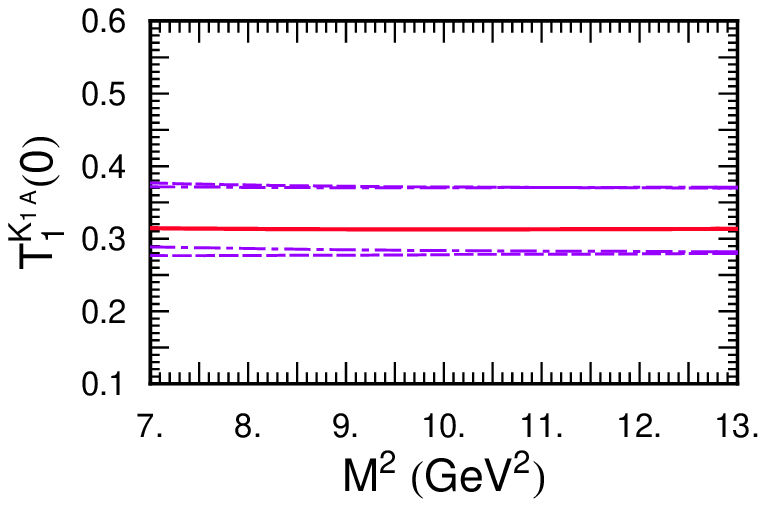}}
   \hskip0.2cm {\epsfxsize=6.8cm\epsffile{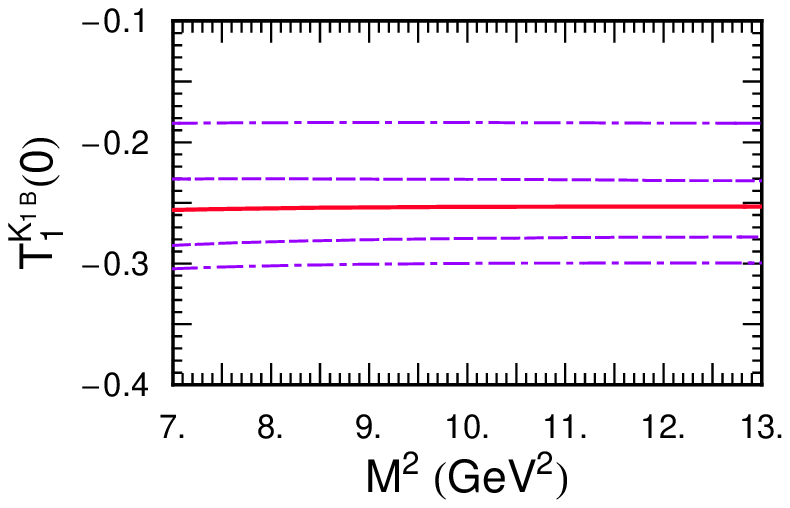}}}
 \vskip0.4cm
\centerline{\parbox{14cm}{\caption{\label{fig:T1} $T_1^{K_{1A}}(0)$ and
$T_1^{K_{1B}}(0)$ as functions of the Borel mass squared, where the central
values of input parameters have been used in the solid curve. The dashed
(dot-dashed) curves are for variation of the $m_{b,pole}$ (parameters for
LCDAs) with the central values of the remaining theoretical parameters.}}}
\end{figure}

In calculating the $B\to K_1 (1270)\gamma$ and $K_1(1400) \gamma$ branching
ratios, $B\to K_1$ tensor form factors have the expressions
 \begin{eqnarray}\label{eq:def-T1}
   T_1^{K_1(1270)}(0) &=& T_1^{K_{1A}}(0)\sin\theta_{K_1}
 + T_1^{K_{1B}}(0)\cos\theta_{K_1}, \nonumber \\
   T_1^{K_1(1400)}(0) &=& T_1^{K_{1A}}(0) \cos\theta_{K_1}
 - T_1^{K_{1B}}(0)\sin\theta_{K_1}.
 \end{eqnarray}
From Eq.~(\ref{eq:T1-SR}), we know that $T_1^{K_{1A}}$ and $T_1^{K_{1B}}$
depend on the definition of the signs of $f_{K_{1A}}$ and $f_{K_{1B}}^\perp$,
so that the resultant $\theta_{K_1}$ also depends on the signs of $f_{K_{1A}}$
and $f_{K_{1B}}^\perp$.

As for the relevant physical properties of $\bar K_{1}$ mesons, we have
  \begin{eqnarray}\label{eq:k1-1v}
 \langle 0 |\bar \psi\gamma_\mu \gamma_5 s |
 \bar K_{1}(1270)(P,\lambda)\rangle
 &=& -i \, f_{K_{1}(1270)}\, m_{K_{1}(1270)}\,\epsilon_\mu^{(\lambda)}
  \nonumber\\
 =-i &{}& \!\!\!\!\!\! (f_{K_{1A}} m_{K_{1A}} \sin{\theta_{K_1}}
    + f_{K_{1B}} m_{K_{1B}} a_0^{\parallel, K_{1B}}
   \cos{\theta_{K_1}})\,\epsilon_\mu^{(\lambda)},~~~~~
 \end{eqnarray}
 \begin{eqnarray}\label{eq:k1-2v}
\langle 0 |\bar \psi\gamma_\mu \gamma_5 s|\bar K_{1}(1400)(P,\lambda)\rangle
  &=& -i \, f_{K_{1}(1400)}\, m_{K_{1}(1400)}\,\epsilon_\mu^{(\lambda)}
  \nonumber\\
 =-i &{}& \!\!\!\!\!\!  (f_{K_{1A}} m_{K_{1A}} \cos {\theta_{K_1}}
     - f_{K_{1B}} m_{K_{1B}} a_0^{\parallel, K_{1B}}
   \sin {\theta_{K_1}})\, \epsilon_\mu^{(\lambda)},~~~~~
 \end{eqnarray}
\begin{eqnarray}\label{eq:k1-1t}
 \langle 0 |\bar \psi\sigma_{\mu\nu}s |\bar K_{1}(1270)(P,\lambda)\rangle
 &=& i f_{K_{1}(1270)}^\perp
 \,\epsilon_{\mu\nu\alpha\beta} \epsilon_{(\lambda)}^\alpha
 P^\beta\nonumber\\
 &=& i (f_{K_{1A}}^\perp a_0^{\perp, K_{1A}}\sin {\theta_K}
    + f_{K_{1B}}^\perp\cos {\theta_K})
 \,\epsilon_{\mu\nu\alpha\beta} \epsilon_{(\lambda)}^\alpha P^\beta,~~~~~
 \end{eqnarray}
and
 \begin{eqnarray}\label{eq:k1-2t}
 \langle 0 |\bar \psi\sigma_{\mu\nu}s |\bar K_{1}(1400)(P,\lambda)\rangle
 &=& i f_{K_{1}(1400)}^\perp
 \,\epsilon_{\mu\nu\alpha\beta} \epsilon_{(\lambda)}^\alpha
 P^\beta\nonumber\\
 &=& i (f_{K_{1A}}^\perp a_0^{\perp, K_{1A}}\cos {\theta_K}
     - f_{K_{1B}}^\perp \sin {\theta_K})
 \,\epsilon_{\mu\nu\alpha\beta}\epsilon_{(\lambda)}^\alpha P^\beta,~~~~~
 \end{eqnarray}
where the values of $f_{K_{1A}}, f_{K_{1B}}^\perp, m_{K_{1A}}, m_{K_{1B}},
a_0^{\parallel, K_{1B}}$ and $a_0^{\perp, K_{1A}}$ are given in Table
\ref{tab:inputs}, and use of  $f_{K_{1B}}=f_{K_{1B}}^\perp(1~{\rm GeV})$ and
$f_{K_{1A}}^\perp=f_{K_{1A}}^\parallel$ is made in the present study. Following
this definition, $a_0^{\parallel, K_{1B}}$ and $a_0^{\perp, K_{1A}}$ vanish in
the SU(3) limit, and we have the relations
\begin{eqnarray}
 \Phi_\perp^{K_1(1270)}(u)
 &=&
 \frac{f_{K_{1A}}^\perp}{f_{K_1(1270)}^\perp}
  \Phi_\perp^{K_{1A}}(u)\sin{\theta_{K_1}}
    + \frac{f_{K_{1B}}^\perp }{f_{K_{1}(1270)}^\perp}
   \Phi_\perp^{K_{1B}}(u)\cos{\theta_{K_1}}, ~~~~\\
  \Phi_\perp^{K_1(1400)}(u)
 &=&
 \frac{f_{K_{1A}}^\perp }{f_{K_1(1400)}^\perp }
  \Phi_\perp^{K_{1A}}(u)\cos{\theta_{K_1}}
    - \frac{f_{K_{1B}}^\perp}{f_{K_1(1400)}^\perp }
   \Phi_\perp^{K_{1B}}(u)\sin{\theta_{K_1}}.
 \end{eqnarray}
In Fig. \ref{fig:Br1} we plot the branching ratios of $B^-\to K_1^-(1270)
\gamma$ and $B^-\to K_1^-(1400) \gamma$ as functions of $\theta_{K_1}$. The
mixing angle dependence of the $K_1^-(1270) \gamma$ mode is opposite to that of
the $K_1^-(1400) \gamma$ mode. To satisfy the observable ${\cal B}(B\to
K_1(1270) \gamma) \gg {\cal B}(B\to K_1 (1400) \gamma)$, we find that the sign
of $\theta_{K_1}$ should be negative. The further constraint for $\theta_{K_1}$
can be obtained from the $\tau^-\to K_1^-(1270)\nu_\tau$ analysis.

\begin{figure}[tb]
 \centerline{{\epsfxsize=8cm \epsffile{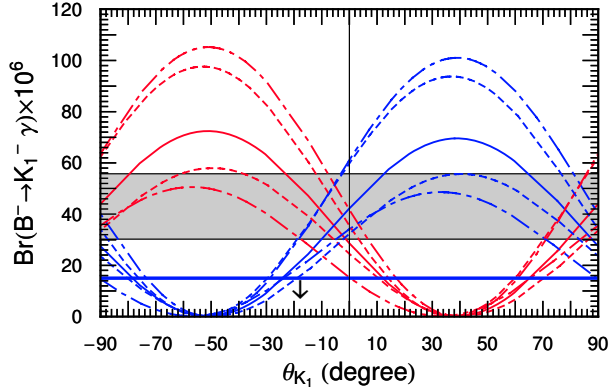}}}
 \vskip0.2cm
 \centerline{\parbox{14cm}{\caption{\label{fig:Br1}
Branching ratios as functions of the mixing angle $\theta_{K_1}$. The upper
five (red) curves at $\theta_{K_1}=-50^\circ$ are for the $K_1(1270)\gamma$
mode, and the lower five (blue) curves for the $K_1(1400)\gamma$ mode. The
solid curves correspond to central values of the input parameters. The
dot-dashed and dashed curves denote the theoretical uncertainties due to the
parameters of LCDAs and $m_{b,pole}$, respectively. The horizontal line is the
experimental limit on $B\to K_1(1400) \gamma$, and the horizontal band shows
the experimental result for the $K_1(1270)\gamma$ mode with its 1$\sigma$
error. }}}
\end{figure}

\subsection{The constraint for $\theta_{K_1}$ from the  $\tau^-\to K_1^-(1270)\nu_\tau$ data}\label{subsec:result-2}

The decay constant $f_{K_1(1270)}$ can be extracted from the measurement
$\tau^-\to K_1^-(1270)\nu_\tau$ by ALEPH \cite{ALEPH}: ${\cal B}(\tau^-\to
K_1^-(1270)\nu_\tau) =(4.7\pm1.1)\times 10^{-3}$, where the formula for the
decay rate is given by
\begin{eqnarray}
 \Gamma(\tau\to K_1\nu_\tau)={G_F^2\over
 16\pi}|V_{us}|^2\,f_{K_1}^2{(m_\tau^2+2m_{K_1}^2)(m_\tau^2-m_{K_1}^2)^2\over
 m_\tau^3}. \label{eq:decay-data-formula}
\end{eqnarray}
It was obtained in  Refs.~\cite{Wang:2007an,ChengDAP} that
\begin{eqnarray}
 \left|f_{K_1(1270)}\right|=169^{+19}_{-21}~{\rm MeV}. \label{eq:decay-data}
\end{eqnarray}
As obtained in the previous subsection, $\theta_{K_1}$ should be negative to
account for the observable ${\cal B}(B\to K_1(1270) \gamma) \gg {\cal B}(B\to
K_1 (1400) \gamma)$. Using the values for $f_{K_{1A}}$ and $f_{K_{1B}}$ as
given in Table~\ref{tab:inputs}, the result for $f_{K_1(1270)}$ in
Eq.~(\ref{eq:decay-data}) and the relation in Eq.~(\ref{eq:k1-1v}), we find
that $a_0^{\parallel, K_{1B}}$ should be negative. Further substituting
$a_0^{\parallel, K_{1B}}=-0.15\pm 0.15$ into Eq.~(\ref{eq:k1-1v}), we obtain
that $\theta_{K_1}$ lies in the interval $-21^\circ\sim -47^\circ$. We can use
the obtained angle to predict the decay constants $f_{K_{1}(1270)}$ and
$f_{K_{1}(1400)}$:
\begin{eqnarray}
f_{K_{1}(1270)} &=& -\left(169^{+25+49}_{-25-40}\right) ~ {\rm MeV}\,, \label{eq:decay1} \\
f_{K_{1}(1400)} &=& ~~~~ 179^{+13+30}_{-13-39}~~~ {\rm MeV}, \label{eq:decay2}
\end{eqnarray}
for $\theta_{K_1}=(-34\pm 13)^\circ$, where the first error is due to the
uncertainties of decay constants and $a_0^{\parallel, K_{1B}}$, and the second
due to the variation of $\theta_{K_1}$. The first error is dominated by the
variation of $a_0^{\parallel, K_{1B}}$. The predicted $\theta_{K_1}=(-34\pm
13)^\circ$ is also consistent with the result given in
Ref.~\cite{Suzuki:1993yc}, where $|\theta_{K_1}|\approx 33^\circ$ or
$57^\circ$. We thus predict
\begin{equation}
{\cal B}(\tau^-\to K_1^-(1400)\nu_\tau) =(3.5^{+0.5+1.2}_{-0.5-1.5})\times 10^{-3},
\end{equation}
to be compared with the current data ${\cal B}(\tau^-\to K_1^-(1400)\nu_\tau)
=(1.7\pm 2.6)\times 10^{-3}$ \cite{PDG} which has large experimental error. If
a more precise measurement for ${\cal B}(\tau^-\to K_1^-(1400)\nu_\tau)$ can
also be achieved, we can extract directly the values of $\theta_{K_1}$ and
$a_0^{\parallel, K_{1B}}$. Consequently, we can have more precise predictions
for the ${\cal B}(B\to K_1(1270) \gamma)$ and ${\cal B}(B\to K_1 (1400)
\gamma)$ branching ratios and $B\to K_1$ transition form factors.

\subsection{$B \to K_1 \gamma$ branching ratios}\label{subsec:result-3}

Using $\overline m_c/\overline m_b=1.25~{\rm GeV}/4.25~{\rm GeV}$, one finds
\begin{eqnarray}
{\cal B} (B \to K_1 \gamma) & = &
\tau_B \,\frac{G_F^2 \alpha |V_{tb} V_{ts}^*|^2}{32 \pi^4} \,
m_{b, {pole}}^2 \, m_B^3 \,
 \left ( 1 - \frac{m_{K_1}^2}{m_B^2} \right )^3 \left(T_1^{K_1}(0)\right)^2
 \nonumber \\
  & & \times
 \left |(-0.360 -i0.015) + A^{(1)K_1}_{\rm sp}(\mu_h) \right|^2,
\label{eq:br-result-r1}
\end{eqnarray}
where $T_1^{K_1(1270)}(0)$ and $T_1^{K_1(1400)}(0)$, as given in
Eq.~(\ref{eq:def-T1}), are $\theta_{K_1}$-dependent. For $\theta_{K_1}=-(34\pm
13)^\circ$, we have
\begin{eqnarray}\label{eq:real-T1s}
T_1^{K_1(1270)}(0) &=&  -\left(0.38^{+0.06+0.08+0.02}_{-0.04-0.07-0.04}\right), \nonumber\\
T_1^{K_1(1400)}(0) &=& ~~~~0.12^{+0.03+0.02+0.08}_{-0.02-0.00-0.09},
\end{eqnarray}
where the first uncertainty comes from the variation of $m_{b,pole}$ and $f_B$
in the sum rules, the second from the parameters of LCDAs, and the third from
$\theta_{K_1}$. To illustrate the contribution due to the hard-spectator
correction, it is interesting to note that, using $\lambda_B=0.35$~GeV,
$\theta_{K_1}=-34^\circ$, $T_1^{K_{1A}}(0)=0.31$, $T_1^{K_{1B}}(0)=-0.25$, and
the center values of the remaining input parameters, we obtain
\begin{eqnarray}
A^{(1)K_1(1270)}_{\rm sp}(\mu_h) &=& 0.016+i0.013, \nonumber\\
A^{(1)K_1(1400)}_{\rm sp}(\mu_h) &=& 0.017-i0.047,
\end{eqnarray}
which suppress the decay rates slightly by about 8\%, in contrast to the $B\to
K^* \gamma$ decay where the interference between the hard-spectator correction
$A^{(1) K^*}_{\rm sp}(\mu_h)=-0.013 -i 0.011$ and the remainder is constructive
\cite{Ali:2001ez}.

In Table~\ref{tab:Brs}, we present a comparison of the resulting branching
ratios in this work with the data. Our results are consistent with the Belle
measurement \cite{Abe:2004kr} within errors.
%
%
\begin{table}[tb]
\renewcommand{\arraystretch}{1.5}
\addtolength{\arraycolsep}{9pt}
$$
\begin{array}{|c c c |}\hline\hline
  & {\cal B}(B^-\to K_1^-(1270)\gamma)
  & {\cal B}(B^-\to K_1^-(1400)\gamma)
 \\ \hline
\begin{array}{c} {\rm Expt.}
              \\ {\rm This\ work}
   \end{array}&
\begin{array}{c} 43\pm13    \\ 66^{+21+30+2+~6}_{-12-24-4-12}
   \end{array}&
\begin{array}{c} <15 \\ 6.5^{+4.0+2.6+0.1+11.9}_{-2.2-0.0-0.2-~5.9}
                  \end{array}
\\ \hline\hline
  & {\cal B}(\bar B^0\to \bar K_1^0(1270)\gamma)
  & {\cal B}(\bar B^0\to \bar K_1^0(1400)\gamma)
 \\ \hline
\begin{array}{c} {\rm Expt.}
              \\ {\rm This\ work}
   \end{array}&
\begin{array}{c} <58   \\  62^{+19+28+2+~5}_{-12-23-4-12}
   \end{array}&
\begin{array}{c} <15   \\ 6.1^{+3.7+2.4+0.0+11.1}_{-2.1-0.0-0.2-~5.5}
                  \end{array}
\\ \hline\hline
\end{array}
$$
\caption{\label{tab:Brs}
Branching ratios for the radiative decays $B \to K_1(1270)\gamma,\,K_1(1400) \gamma$
(in units of $10^{-6}$) in this work and the experiment \cite{Abe:2004kr}.
The branching ratios correspond to $\theta_{K_1}=-(34^\circ\pm 13^\circ)$
in our work, where the first error comes from the variation of $m_{b,pole}$
and $f_B$, the second from the parameters of LCDAs, the third from $\lambda_B$,
and the forth from $\theta_{K_1}$. The annihilation amplitudes are not included
in the neutral $B$ decay modes.}
\end{table}
A much more precise determination of $\theta_{K_1}$ can be made by the measurement
\begin{equation}\label{eq:R}
R_{K_1}=\frac{{\cal B}(B\to K(1400)\gamma)}{{\cal B}(B\to K(1270)\gamma)}.
\end{equation}
The current upper bound of this ratio is $R_{K_1}<0.5$. It can be seen from
Fig.~\ref{fig:Br1-ratio} that $R_{K_1}$ weakly depends on the theoretical
uncertainty. Thus, $R_{K_1}$ is a suitable quantity for measuring the mixing
angle $\theta_{K_1}$. In the light-cone sum rule calculation, the physical
quantities, including the branching ratios and transition form factors, receive
large errors from the uncertainties of G-parity violating Gegenbaur moments. A
more precise value for $\theta_{K_1}$ can be used to extract a better result of
$a_0^{\parallel, K_{1B}}$ from the data for ${\cal B}(\tau^-\to
K_1^-(1270)\nu_\tau)$; the remaining G-parity violating Gegenbaur moments can
thus be determined using Eq.~(141) in Ref.~\cite{Yang:2007zt}. On the other
hand, we can also obtain good estimates for $\theta_{K_1}$ and $a_0^{\parallel,
K_{1B}}$ from the data ${\cal B}(\tau^-\to K_1^-(1270)\nu_\tau)$ and ${\cal
B}(\tau^-\to K_1^-(1400)\nu_\tau)$ if we can improve the measurement for ${\cal
B}(\tau^-\to K_1^-(1400)\nu_\tau)$. Consequently, theoretical uncertainties due
to G-parity violating Gegenbaur moments and $\theta_{K_1}$ can be reduced in
the form factors and branching ratios calculations.

\begin{figure}[tb]
 \centerline{{\epsfxsize=8cm\epsffile{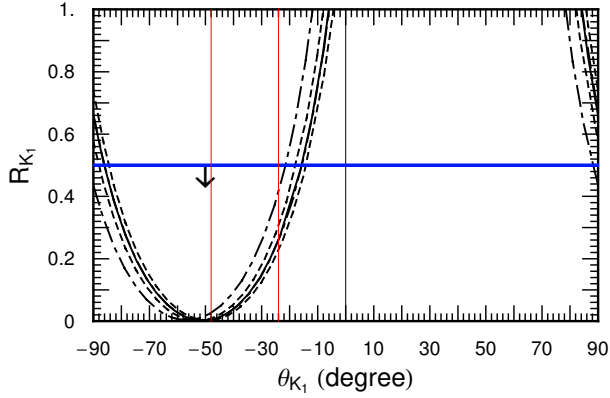}}}
 \vskip0.2cm
\centerline{\parbox{14cm}{\caption{\label{fig:Br1-ratio} Same as
Fig.~\ref{fig:Br1} except for the ratio $R_{K_1}={\cal B}(B\to
K_1(1400)\gamma)/{\cal B}(B\to K_1(1270)\gamma)$ as a function of the mixing
angle $\theta_{K_1}$. }}}
\end{figure}
%

\section{Conclusions}\label{sec:conclusion}

We have presented a detailed study of $B\to K_1(1270) \gamma$ and $B\to
K_1(1400) \gamma$ decays. Our main results are as follows.

\begin{itemize}

\item
Using the light-cone sum rule technique, we have evaluated the $B\to K_{1A},
K_{1B}$ tensor form factors, $T_1^{K_{1A}}(0)$ and $T_1^{K_{1B}}(0)$, where the
contributions have been included up to the first order in $m_{K_1}/m_b$. We
obtain $T_1^{K_{1A}}(0) =0.31^{+0.06+0.01+0.06}_{-0.04-0.01-0.03}$ and
$T_1^{K_{1B}}(0) = - (0.25_{-0.02-0.01-0.07}^{+0.03+0.01+0.05})$.

\item
The sign ambiguity of the  $K_1(1270)$--$K_1(1400)$ mixing angle $\theta_{K_1}$
can be resolved by defining $f_{K_{1A}}$ and $f_{K_{1B}}^\perp$ to be positive.
Combining the analysis for the decays $B\to K_1 \gamma$ and $\tau^-\to
K_1^-(1270)\nu_\tau$, we find that the mixing angle $\theta_{K_1}$ should be
negative, and its value lies in the interval $-(34 \pm 13)^\circ$.  We obtain
$f_{K_{1}(1270)} = -\left(169^{+25+49}_{-25-40}\right)$ MeV and
$f_{K_{1}(1400)} = 179^{+13+30}_{-13-39}$ MeV, and predict $ {\cal B}(\tau^-\to
K_1^-(1400)\nu_\tau) =(3.5^{+0.5+1.2}_{-0.5-1.5})\times 10^{-3}.$

\item
We find $T_1^{K_1(1270)}(0) = - (0.38^{+0.06+0.08+0.02}_{-0.04-0.07-0.04} ),
T_1^{K_1(1400)}(0)  = 0.12^{+0.03+0.02+0.08}_{-0.02-0.00-0.09}$. The
hard-spectator contribution suppresses the  $B\to K_1(1270) \gamma$ and $B\to
K_1(1400) \gamma$ decay rates slightly by about 8\%, in contrast with the
situation for $B\to K^*\gamma$. The predicted branching ratios for the decays
$B\to K_1(1270) \gamma$ and $B\to K_1(1400) \gamma$ agree with the data within
the errors.

\item
We point out that better determinations of the $\theta_{K_1}$ and G-parity
violating Gegenbaur moments of leading-twist light-cone distribution amplitudes
can be obtained from a more precise measurement for the ratio $R_{K_1}={\cal
B}(B\to K_1(1400)\gamma)/{\cal B}(B\to K_1(1270)\gamma)$ or from an improved
measurement for  ${\cal B}(\tau^-\to K_1^-(1400)\nu_\tau)$ together with the
${\cal B}(\tau^-\to K_1^-(1270)\nu_\tau)$ data. Thus, the theoretical
uncertainties can be further reduced.

\end{itemize}

\section*{Acknowledgments}

This research was supported in part by the National Science Council of R.O.C. under
Grant No. NSC96-2112-M-033-004-MY3 and No. NSC96-2811-M-033-004.

\vskip1cm

\appendix
\section{Two-parton distribution amplitudes}\label{app:2da-def}

In the calculation, the LCDAs of the axial-meson appear in the
following way
 \begin{eqnarray}
  &&  \langle \bar K_1(P,\lambda)|\bar s_{\alpha}(y) \, \psi_{\delta}(x)|0\rangle
= -\frac{i}{4} \, \int_0^1 du \,  e^{i (u \, P y +
    \bar u P x)}\,\Bigg\{ f_{K_1} m_{K_1} \Bigg[
  \not\! P\gamma_5 \, \frac{\epsilon^*_{(\lambda)} z}{Pz} \, \Phi_\parallel(u)
\nonumber\\[0.1cm]
  && \qquad\,\,\,
  +\Bigg( \not\! \epsilon^* -\not\! P \frac{\epsilon^*_{(\lambda)} z}{Pz}\Bigg)\,
   \gamma_5 g_\perp^{(a)}(u)  - \not\! z\gamma_5 \frac{\epsilon^*_{(\lambda)} z}{2(Pz)^2}
  m_{K_1}^2 \bar g_3(u) +
 \epsilon_{\mu\nu\rho\sigma} \,
    \epsilon^*_{(\lambda)}{}^\nu  p^{\rho} z^\sigma \, \gamma^\mu
    \, \frac{g_\perp^{(v)}(u)}{4}\Bigg]
\nonumber \\[0.1em]
  && \qquad\,\,\,
  + \,f^{\perp}_{K_1} \Bigg[
  \frac{1}{2}\bigg( \! \not\! P\not\!\epsilon^*_{(\lambda)}-
  \not\!\epsilon^*_{(\lambda)} \not\! P  \bigg) \gamma_5\Phi_\perp(u)
 -
   \frac{1}{2}\bigg( \! \not\! P\not\! z- \not\! z \not\! P  \bigg)
   \gamma_5 \frac{\epsilon^*_{(\lambda)} z}{(Pz)^2} m_{K_1}^2 \bar
   h_\parallel^{(t)} (u)\nonumber\\
 && \qquad
 \,\,\,
 -\frac{1}{4}\bigg( \! \not\!\epsilon^*_{(\lambda)} \not z-
  \not z \not\!\epsilon^*_{(\lambda)} \bigg) \gamma_5
  \frac{m_{K_1}^2 }{Pz} \bar h_3 (u) +i \big(\epsilon^*_{(\lambda)} z\big) m_{K_1}^2
 \gamma_5
 \frac{h^{(p)}_\parallel (u)}{2}
  \Bigg]\Bigg\}_{\delta\alpha} +{\cal O}\Big((x-y)^2\Big)\,,\nonumber\\
 \end{eqnarray}
where
\begin{eqnarray}
 \bar g_3 (u) &=& g_3(u) +\Phi_\parallel -2 g_\perp^{(a)}(u),\nonumber\\
 \bar h_\parallel^{(t)}(u) &=& h_\parallel^{(t)}(u)
   - \frac{1}{2} \Phi_\perp(u) -\frac{1}{2} h_3(u) , \nonumber\\
 \bar h_3(u) &=& h_3(u) -\Phi_\perp(u),
\end{eqnarray}
$z^2=(y-x)^2 \neq 0$, and ${P}^2=m_{K_1}^2$. The detailed LCDAs are defined in
Ref.~\cite{Yang:2007zt}. Here $\Phi_\parallel, \Phi_\perp$ are of twist-2,
$g_\perp^{(a)}, g_\perp^{(v)}, h_\parallel^{(t)}, h_\parallel^{(p)}$ of
twist-3, and $g_3, h_3$ of twist-4. In SU(3) limit, due to G-parity,
$\Phi_\parallel, g_\perp^{(a)}$, $g_\perp^{(v)}$, and $g_3$ are symmetric
(antisymmetric) under the replacement $u\leftrightarrow 1-u$ for the $1^3P_1$
($1^1P_1$) states, whereas $\Phi_\perp, h_\parallel^{(t)}$,
$h_\parallel^{(p)}$, and $h_3$ are antisymmetric (symmetric). For convenience,
we normalize the distribution amplitudes of the $1^3P_1$ and $1^1P_1$ states to
be subject to
 \begin{equation}
 \int_0^1 du\Phi_\parallel(u)=1, \ \ \
  \int_0^1 du\Phi_\perp(u)=1 .
 \end{equation}
We take $f_{^3\! P_1}^\perp=f_{^3\! P_1}$ and $f_{^1\! P_1}=f_{^1\!
P_1}^\perp(\mu=1~{\rm GeV})$ in the study, such that we define
 \begin{eqnarray}
 &&\langle \bar K_{1A}(P,\lambda)|
  \bar s(0) \sigma_{\mu\nu}\gamma_5 \psi(0)
   |0\rangle
  =  f_{K_{1A}}^{\perp} a_0^{\perp,K_{1A}} \,
(\epsilon^{*(\lambda)}_{\mu} P_{\nu} - \epsilon_{\nu}^{*(\lambda)} \nonumber
P_{\mu}),
 \\
 &&  \langle \bar K_{1B}(P,\lambda)|\bar s(0) \gamma_\mu \gamma_5 \psi(0)|0\rangle
   = if_{K_{1B}} a_0^{\parallel,K_{1B}} \, m_{K_{1B}} \,  \epsilon^{*(\lambda)}_\mu
   \,,\label{app:decay}
 \end{eqnarray}
where $a_0^{\perp,K_{1A}}$ and  $a_0^{\parallel,K_{1B}}$ are the
Gegenbauer zeroth moments, which vanish in the SU(3) limit.

We take into account the approximate forms of twist-2 distributions
for the $\bar K_{1A}$ meson to be \cite{Yang:2007zt}
\begin{eqnarray}
\Phi_\parallel(u) & = & 6 u \bar u \left[ 1 + 3 a_1^\parallel\, \xi +
a_2^\parallel\, \frac{3}{2} ( 5\xi^2  - 1 )
 \right], \label{eq:lcda-3p1-t2-1}\\
 \Phi_\perp(u) & = & 6 u \bar u \left[ a_0^\perp + 3 a_1^\perp\, \xi +
a_2^\perp\, \frac{3}{2} ( 5\xi^2  - 1 ) \right], \label{eq:lcda-3p1-t2-2}
\end{eqnarray}
and for the $\bar K_{1B}$ meson to be
\begin{eqnarray}
 \Phi_\parallel(u) & = & 6 u \bar u \left[ a_0^\parallel + 3
a_1^\parallel\, \xi +
a_2^\parallel\, \frac{3}{2} ( 5\xi^2  - 1 ) \right], \label{eq:lcda-1p1-t2-1}\\
\Phi_\perp(u) & = & 6 u \bar u \left[ 1 + 3 a_1^\perp\, \xi +
a_2^\perp\, \frac{3}{2} ( 5\xi^2  - 1 ) \right],
\label{eq:lcda-1p1-t2-2}
\end{eqnarray}
where $\xi=2u-1$.

For the two-parton twist-3 chiral-even LCDAs, which are relevant here, we take
the approximate expressions up to conformal spin $9/2$ and ${\cal O}(m_s)$
\cite{Yang:2007zt}:
\begin{eqnarray}
 g_\perp^{(a)}(u) & = &  \frac{3}{4}(1+\xi^2)
+ \frac{3}{2}\, a_1^\parallel\, \xi^3
 + \left(\frac{3}{7} \,
a_2^\parallel + 5 \zeta_{3,K_{1A}}^V \right) \left(3\xi^2-1\right)
 \nonumber\\
& & {}+ \left( \frac{9}{112}\, a_2^\parallel + \frac{105}{16}\,
 \zeta_{3,K_{1A}}^A - \frac{15}{64}\, \zeta_{3,K_{1A}}^V \omega_{K_{1A}}^V
 \right) \left( 35\xi^4 - 30 \xi^2 + 3\right) \nonumber\\
 & &
 + 5\Bigg[ \frac{21}{4}\zeta_{3,K_{1A}}^V \sigma_{K_{1A}}^V
  + \zeta_{3,K_{1A}}^A \bigg(\lambda_{K_{1A}}^A -\frac{3}{16}
 \sigma_{K_{1A}}^A\Bigg) \Bigg]\xi(5\xi^2-3)
 \nonumber\\
& & {}-\frac{9}{2} \bar{a}_1^\perp
\,\widetilde{\delta}_+\,\left(\frac{3}{2}+\frac{3}{2}\xi^2+\ln u
 +\ln\bar{u}\right) - \frac{9}{2} \bar{a}_1^\perp\,\widetilde{\delta}_-\, (
3\xi + \ln\bar{u} - \ln u), \label{eq:ga-3p1}\\
g_\perp^{(v)}(u) & = & 6 u \bar u \Bigg\{ 1 +
 \Bigg(a_1^\parallel + \frac{20}{3} \zeta_{3,K_{1A}}^A
 \lambda_{K_{1A}}^A\Bigg) \xi\nonumber\\
 && + \Bigg[\frac{1}{4}a_2^\parallel + \frac{5}{3}\,
 \zeta^V_{3,K_{1A}} \left(1-\frac{3}{16}\, \omega^V_{K_{1A}}\right)
 +\frac{35}{4} \zeta^A_{3,K_{1A}}\Bigg] (5\xi^2-1) \nonumber\\
 &&+ \frac{35}{4}\Bigg(\zeta_{3,K_{1A}}^V
 \sigma_{K_{1A}}^V -\frac{1}{28}\zeta_{3,K_{1A}}^A
 \sigma_{K_{1A}}^A \Bigg) \xi(7\xi^2-3) \Bigg\}\nonumber\\
& & {} -18 \, a_1^\perp\widetilde{\delta}_+ \,  (3u \bar{u} +
\bar{u} \ln \bar{u} + u \ln u ) - 18\,
a_1^\perp\widetilde{\delta}_- \,  (u \bar u\xi + \bar{u} \ln \bar{u} -
u \ln u),
 \label{eq:gv-3p1}
 \end{eqnarray}
for the $\bar K_{1A}$ state, and
\begin{eqnarray}
 g_\perp^{(a)}(u) & = & \frac{3}{4} a_0^\parallel (1+\xi^2)
+ \frac{3}{2}\, a_1^\parallel\, \xi^3
 + 5\left[\frac{21}{4} \,\zeta_{3,K_{1B}}^V
 + \zeta_{3,K_{1B}}^A \Bigg(1-\frac{3}{16}\omega_{K_{1B}}^A\Bigg)\right]
 \xi\left(5\xi^2-3\right)
 \nonumber\\
& & {}+ \frac{3}{16}\, a_2^\parallel \left(15\xi^4 -6 \xi^2 -1\right)
 + 5\, \zeta^V_{3,K_{1B}}\lambda^V_{K_{1B}}\left(3\xi^2 -1\right)
 \nonumber\\
& & {}+ \frac{105}{16}\left(\zeta^A_{3,K_{1B}}\sigma^A_{K_{1B}}
-\frac{1}{28} \zeta^V_{K_{1B}}\sigma^V_{K_{1B}}\right)
 \left(35\xi^4 -30 \xi^2 +3\right)\nonumber\\
 & & {}-15\bar{a}_2^\perp \bigg[ \widetilde{\delta}_+ \xi^3 +
 \frac{1}{2}\widetilde{\delta}_-(3\xi^2-1) \bigg] \nonumber\\
& & {}
  -\frac{3}{2}\,\bigg[\widetilde{\delta}_+\, ( 2 \xi + \ln\bar{u} -\ln u)
 +\, \widetilde{\delta}_-\,(2+\ln u + \ln\bar{u})\bigg](1+6a_2^\perp)
 ,\label{eq:ga-1p1}\\
g_\perp^{(v)}(u) & = & 6 u \bar u \Bigg\{ a_0^\parallel +
a_1^\parallel \xi +
 \Bigg[\frac{1}{4}a_2^\parallel
  +\frac{5}{3} \zeta^V_{3,K_{1B}}
  \Bigg(\lambda^V_{K_{1B}} -\frac{3}{16} \sigma^V_{K_{1B}}\Bigg)
  +\frac{35}{4} \zeta^A_{3,K_{1B}}\sigma^A_{K_{1B}}\Bigg](5\xi^2-1) \nonumber\\
  & & {}  + \frac{20}{3}\,  \xi
 \left[\zeta^A_{3, K_{1B}}
 + \frac{21}{16}
 \Bigg(\zeta^V_{3,K_{1B}}- \frac{1}{28}\, \zeta^A_{3,K_{1B}}\omega^A_{K_{1B}}
  \Bigg)
 (7\xi^2-3)\right]\nonumber\\
 & & {} -5\, a_2^\perp [2\widetilde\delta_+ \xi + \widetilde\delta_- (1+\xi^2)]
 \Bigg\}\nonumber\\
 & & {} - 6 \bigg[\, \widetilde{\delta}_+ \, (\bar{u} \ln\bar{u} -u\ln u )
  +\, \widetilde{\delta}_- \, (2u \bar{u} + \bar{u} \ln \bar{u} + u \ln u)\bigg]
  (1+6 a_2^\perp) ,
 \label{eq:gv-1p1}
\end{eqnarray}
for the $\bar K_{1B}$ state, where
\begin{equation}
\widetilde{\delta}_\pm  =\pm {f_{K_1}^{\perp}\over f_{K_1}}{m_{s}
\over m_{K_1}},\qquad \zeta_{3,K_1}^{V,A} = \frac{f^{V,A}_{3K_1}}{f_{K_1} m_{K_1}}.
\label{eq:parameters3}
\end{equation}

\section{Three-parton chiral-even distribution amplitudes of twist-3}\label{app:3da-def}

Taking into account the contributions up to terms of conformal spin $9/2$ and
considering the corrections of order $m_s$, the twist-3 three-parton
chiral-even distribution amplitudes, defined in Eqs.~(\ref{eq:3plcda-t3-1}) and
(\ref{eq:3plcda-t3-2}), can be approximately written as \cite{Yang:2007zt}
 \begin{eqnarray}
  {\cal A} (\underline{\alpha}) &=&5040 (\alpha_{s}-\alpha_{\psi})\alpha_{s}\alpha_{\psi}\alpha_{g}^2
  +360\alpha_{s}\alpha_{\psi}\alpha_{g}^2
  \Big[ \lambda^A_{K_{1A}}+ \sigma^A_{K_{1A}}\frac{1}{2}(7\alpha_g-3)\Big],
 \label{eq:lcda-3p1-t3-1}\\
  {\cal V}(\underline{\alpha}) &=&360\alpha_{s}\alpha_{\psi}\alpha_{g}^2
  \Big[ 1+ \omega^V_{K_{1A}}\frac{1}{2}(7\alpha_g-3)\Big]
 +5040 (\alpha_{s}-\alpha_{\psi})\alpha_{s}\alpha_{\psi}\alpha_{g}^2
 \sigma^V_{K_{1A}},\label{eq:lcda-3p1-t3-2}
 \end{eqnarray}
for the $\bar K_{1A}$ state, and
 \begin{eqnarray}
  {\cal A}(\underline{\alpha}) &=&360 \alpha_{s}\alpha_{\psi}\alpha_{g}^2
  \Big[ 1+ \omega^A_{K_{1B}}\frac{1}{2}(7\alpha_g-3)\Big]
  +5040 (\alpha_{s}-\alpha_{\psi})\alpha_{s}\alpha_{\psi}\alpha_{g}^2
 \sigma^A_{K_{1B}}, \label{eq:lcda-1p1-t3-1}\\
  {\cal V}(\underline{\alpha}) &=&5040 (\alpha_{s}-\alpha_{\psi})\alpha_{s}\alpha_{\psi}\alpha_{g}^2
  +360\alpha_{s}\alpha_{\psi}\alpha_{g}^2
  \Big[ \lambda^V_{K_{1B}}+ \sigma^V_{K_{1B}}\frac{1}{2}(7\alpha_g-3)\Big],
  \label{eq:lcda-1p1-t3-2}
 \end{eqnarray}
for the $\bar K_{1B}$ state, where $\lambda$'s correspond to conformal
spin 7/2, while $\omega$'s and $\sigma$'s are parameters with
conformal spin 9/2. Note that as the SU(3)-symmetry (and G-parity) is
restored, we have $\lambda$'s=$\sigma$'s=0.



\begin{thebibliography}{99}

\bibitem{Coan:1999kh}
  T.~E.~Coan {\it et al.}  [CLEO Collaboration],
  Phys.\ Rev.\ Lett.\  {\bf 84}, 5283 (2000)
  [arXiv:hep-ex/9912057].

\bibitem{Nakao:2004th}
  M.~Nakao {\it et al.}  [BELLE Collaboration],
  Phys.\ Rev.\  D {\bf 69}, 112001 (2004)
  [arXiv:hep-ex/0402042].

\bibitem{Aubert:2004te}
  B.~Aubert {\it et al.}  [BABAR Collaboration],
  Phys.\ Rev.\  D {\bf 70}, 112006 (2004)
  [arXiv:hep-ex/0407003].


\bibitem{Abe:2004kr}
  K.~Abe {\it et al.}  [BELLE Collaboration],
  arXiv:hep-ex/0408138;
  H.~Yang {\it et al.},
  Phys.\ Rev.\ Lett.\  {\bf 94}, 111802 (2005)
  [arXiv:hep-ex/0412039];
 Heavy Flavor Averaging Group, E. Barberio {\it et al.,}
arXiv:0704.3575 [hep-ex] and online update at
 http://www.slac.stanford.edu/xorg/hfag.


\bibitem{Altomari:1987pv}
  T.~Altomari,
  Phys.\ Rev.\  D {\bf 37}, 677 (1988).

\bibitem{Veseli:1995bt}
  S.~Veseli and M.~G.~Olsson,
  Phys.\ Lett.\  B {\bf 367}, 309 (1996)
  [arXiv:hep-ph/9508255].

\bibitem{Safir:2001cd}
  A.~S.~Safir,
  Eur.\ Phys.\ J.\ directC {\bf 3} (2001) 1
  [arXiv:hep-ph/0109232].

\bibitem{Cheng:2004yj}
  H.~Y.~Cheng and C.~K.~Chua,
  Phys.\ Rev.\  D {\bf 69}, 094007 (2004)
  [arXiv:hep-ph/0401141].

\bibitem{Lee:2004ju}
  J.~P.~Lee,
  Phys.\ Rev.\  D {\bf 69}, 114007 (2004)
  [arXiv:hep-ph/0403034].

\bibitem{Kwon:2004ri}
  Y.~J.~Kwon and J.~P.~Lee,
  Phys.\ Rev.\  D {\bf 71}, 014009 (2005)
  [arXiv:hep-ph/0409133].


\bibitem{PDG} Particle Data Group, Y.M. Yao {\it et al.,} J. Phys. G
{\bf 33}, 1 (2006).

\bibitem{BaBara1pi}
  B.~Aubert {\it et al.}  (BaBar Collaboration),
  Phys.\ Rev.\ Lett.\  {\bf 97}, 051802 (2006).

\bibitem{Bellea1pi} K. Abe {\it  et al.} (Belle Collaboration),
arXiv:0706.3279 [hep-ex].

\bibitem{Aubert:2007xd}
  B.~Aubert {\it et al.}  [The BABAR Collaboration],
  Phys.\ Rev.\ Lett.\  {\bf 99}, 241803 (2007)
  [arXiv:0707.4561 [hep-ex]].


\bibitem{:2007kp}
  B.~Aubert {\it et al.}  [BABAR Collaboration],
  Phys.\ Rev.\ Lett.\  {\bf 99}, 261801 (2007)
  [arXiv:0708.0050 [hep-ex]].

\bibitem{Aubert:2007ds}
  B.~Aubert {\it et al.}  [BABAR Collaboration],
  Phys.\ Rev.\ Lett.\  {\bf 100}, 051803 (2008)
  [arXiv:0709.4165 [hep-ex]].

\bibitem{Yang1P1} K.C. Yang,  Phys.\ Rev.\  D {\bf 72}, 034009 (2005);
D {\bf 72}, (E)059901 (2005).

\bibitem{Chen05} C.H. Chen, C.Q. Geng, Y.K. Hsiao, and Z.T. Wei, \pr D
{\bf 72}, 054011 (2005).

\bibitem{Nardulli05} G. Nardulli and T.N. Pham, \pl B {\bf 623}, 65
(2005).

\bibitem{Nardulli07}
  V.~Laporta, G.~Nardulli and T.~N.~Pham,
  Phys.\ Rev.\  D {\bf 74}, 054035 (2006)
  [Erratum-ibid.\  D {\bf 76}, 079903 (2007)]
  [arXiv:hep-ph/0602243].

\bibitem{Calderon}
  G.~Calderon, J.~H.~Munoz and C.~E.~Vera,
  Phys.\ Rev.\  D {\bf 76}, 094019 (2007)
  [arXiv:0705.1181 [hep-ph]].

\bibitem{Yang:2007sb}
  K.~C.~Yang,
  Phys.\ Rev.\  D {\bf 76}, 094002 (2007)
  [arXiv:0705.4029 [hep-ph]].

\bibitem{Cheng:2007mx}
  H.~Y.~Cheng and K.~C.~Yang,
  Phys.\ Rev.\  D {\bf 76}, 114020 (2007)
  [arXiv:0709.0137 [hep-ph]].

\bibitem{Suzuki:1993yc}
  M.~Suzuki,
  Phys.\ Rev.\  D {\bf 47} (1993) 1252.

\bibitem{Burakovsky:1997ci}
  L.~Burakovsky and J.~T.~Goldman,
  Phys.\ Rev.\  D {\bf 57}, 2879 (1998)
  [arXiv:hep-ph/9703271].

\bibitem{ChengDAP} H.Y. Cheng, \pr D {\bf 67}, 094007 (2003).

\bibitem{Yang:2007zt}
  K.~C.~Yang,
  Nucl. Phys. B {\bf 776}, 187 (2007)
  [arXiv:0705.0692 [hep-ph]].

\bibitem{Yang:2005gk}
  K.~C.~Yang,
  JHEP {\bf 0510}, 108 (2005)
  [arXiv:hep-ph/0509337].

\bibitem{Lee:2006qj}
  J.~P.~Lee,
  Phys.\ Rev.\  D {\bf 74}, 074001 (2006)
  [arXiv:hep-ph/0608087].

\bibitem{Wang:2007an}
  W.~Wang, R.~H.~Li and C.~D.~Lu,
  arXiv:0711.0432 [hep-ph].

\bibitem{Balitsky:1989ry}
  I.~I.~Balitsky, V.~M.~Braun and A.~V.~Kolesnichenko,
  Nucl.\ Phys.\  B {\bf 312}, 509 (1989).

\bibitem{Chernyak:1990ag}
  V.~L.~Chernyak and I.~R.~Zhitnitsky,
  Nucl.\ Phys.\  B {\bf 345} (1990) 137.

\bibitem{Belyaev:1993wp}
  V.~M.~Belyaev, A.~Khodjamirian and R.~Ruckl,
  Z.\ Phys.\  C {\bf 60}, 349 (1993)
  [arXiv:hep-ph/9305348].

\bibitem{Ball:1997rj}
  P.~Ball and V.~M.~Braun,
  Phys.\ Rev.\  D {\bf 55}, 5561 (1997)
  [arXiv:hep-ph/9701238].

\bibitem{Ball:2004rg}
 See, for example, P.~Ball and R.~Zwicky,
  Phys.\ Rev.\  D {\bf 71}, 014029 (2005)
  [arXiv:hep-ph/0412079], and references therein.

\bibitem{Greub:1996tg}
  C.~Greub, T.~Hurth and D.~Wyler,
  Phys.\ Rev.\  D {\bf 54}, 3350 (1996)
  [arXiv:hep-ph/9603404].

\bibitem{Ali:2001ez}
 M.~Beneke, T.~Feldmann and D.~Seidel,
  Nucl.\ Phys.\  B {\bf 612} (2001) 25
  [arXiv:hep-ph/0106067];
 A.~Ali and A.~Y.~Parkhomenko,
  Eur.\ Phys.\ J.\  C {\bf 23}, 89 (2002)
  [arXiv:hep-ph/0105302].

\bibitem{Bosch:2001gv}
  S.~W.~Bosch and G.~Buchalla,
  Nucl.\ Phys.\  B {\bf 621}, 459 (2002)
  [arXiv:hep-ph/0106081].

\bibitem{Beneke:2000ry}
  M.~Beneke, G.~Buchalla, M.~Neubert and C.~T.~Sachrajda,
  Nucl.\ Phys.\  B {\bf 591}, 313 (2000)
  [arXiv:hep-ph/0006124].

\bibitem{Beneke:2000wa}
  M.~Beneke and T.~Feldmann,
  Nucl.\ Phys.\  B {\bf 592}, 3 (2001)
  [arXiv:hep-ph/0008255].

\bibitem{Ball:2006eu}
  P.~Ball, G.~W.~Jones and R.~Zwicky,
  Phys.\ Rev.\  D {\bf 75}, 054004 (2007)
  [arXiv:hep-ph/0612081].

\bibitem{Becirevic:2006nm}
  D.~Becirevic, V.~Lubicz and F.~Mescia,
  Nucl.\ Phys.\  B {\bf 769}, 31 (2007)
  [arXiv:hep-ph/0611295].

\bibitem{ALEPH} R. Barate {\it et al.} (ALEPH Collaboration),  Eur. J.
Phys. C {\bf 11}, 599 (1999).

\end{thebibliography}
 \end{document}